\documentclass[11pt]{article}
\usepackage{cite}
\usepackage{graphicx}
\usepackage{amsmath}
\usepackage{authblk}

\newcommand{\WpWm} {W^+W^-}
\newcommand{\ttbar}{t\bar{t}}
\newcommand{\MET}{E_T^{\mathrm{miss}}}
\newcommand{\METproj}{E_{T\,{\mathrm{proj}}}^{\mathrm{miss}}}
\newcommand{\MLL}{M_{\ell\ell}}
\newcommand{\csWW}{\sigma_{WW}}
\newcommand{\TeV}{{\mathrm{TeV}}}
\newcommand{\GeV}{{\mathrm{GeV}}}
\newcommand{\opencv}{{\sc OpenCV}}
\newcommand{\pythia}{{\sc Pythia}}
\newcommand{\pyclus}{{\tt PYCLUS}}
\newcommand{\Nextra}{N_{\ell}^{\mathrm{extra}}}
\newcommand{\pTextra}{p_T^{\mathrm{extra}}}
\newcommand{\Njet}{N_{\mathrm{jet}}}
\newcommand{\Njetb}{N_{b-\mathrm{jet}}}
\newcommand{\pTjeti}{p^{\mathrm{jet}}_{Ti}}
\newcommand{\LChan}{F_{\ell\ell}}
\newcommand{\Ycut}{Y_{\mathrm{cut}}}
\newcommand{\effWW}{\epsilon_{WW}}
\newcommand{\efftt}{\epsilon_{tt}}
\newcommand{\AWW}{A_{WW}}
\newcommand{\Att}{A_{tt}}
\newcommand{\sigmaWW}{\sigma_{WW}}
\newcommand{\sigmatt}{\sigma_{tt}}
\newcommand{\Lum} {{\cal{L}}}
\newcommand{\NWW}{N_{WW}}
\newcommand{\Ntt}{N_{tt}}
\newcommand{\pb}{{\mathrm{pb}}}

\title{\bf Using Random Forests to Classify $\WpWm$ and $\ttbar$ Events}
\author[1]{J. Lovelace Rainbolt}
\author[1]{Thoth Gunter}
\author[1]{Michael Schmitt}
\affil[1]{Department of Physics and Astronomy, Northwestern University, Evanston, IL 60091}
\begin{document}
\maketitle
\section{Introduction}
\par
The Large Hadron Collider at CERN collides proton beams at a center-of-mass energy 
of~$7$--$8~\TeV$, and the experiments arrayed there gather immense
data sets rich with information based on reconstructed particles and
other objects.  In this context, an ``event'' corresponds to a beam crossing
in which one or more pairs of protons collide and produce dozens or even
hundreds of particles.  Physicists can identify many ``features'' in these events
that allow them to be classified.   For example, an event with a pair of
oppositely-charged energetic leptons (electrons or muons), missing transverse
energy, and a pair of jets could be selected as a candidate $pp \rightarrow \ttbar$
event, where ``$t$'' stands for a top quark.
Event classification is essential for measuring cross sections and also
in the search for new particles and phenomena.
\par
Measurements of the cross section $\csWW$ for the process $pp \rightarrow \WpWm$ provide
important tests of the standard model (SM) of particle physics.  
The first measurements~\cite{CMSWW,ATLASWW}
focus on the case in which both $W$ bosons decay leptonically,
in which case $\WpWm$ events are characterized by two features
shared with $\ttbar$ events: two 
oppositely-charged energetic leptons and missing transverse energy.
The number of reconstructed jets allows
one to distinguish $\WpWm$ from $\ttbar$ events on a statistical basis;
the $\ttbar$ ``background'' to the $\WpWm$ ``signal'' is nearly eliminated
by rejecting events with any jets, but with a significant reduction of the signal.
Currently, the measured values for $\csWW$ are larger than predicted
by advanced theoretical calculations (see the Appendix).
It has been suggested that the requirement
of zero jets (known as the ``jet veto'') is responsible for the 
discrepancy~\cite{Meade,Jaiswal,Monni}.
\par
We have conducted an exercise in supervised machine learning, in which
we use a Random Forest~(RF) classifier to separate events from the two similar
processes $pp \rightarrow \WpWm$ and $pp \rightarrow \ttbar$.
RF classifiers have been used in particles physics by the
D\O\ Collabration~\cite{D0top,D0VV,D0Higgs,D0bjet} and
by the BaBar Collaboration~\cite{BaBarbsgamma,BaBarHiggs}.
Our exercise is a step toward a broader application of RF classifiers and
is meant only to explore the RF classifier in a specific and well-defined case.
The results are compared to a version of the original analysis published
by the CMS~Collaboration~\cite{CMSWW}.  We find that the RF classifier
provides a significantly more powerful rejection of $\ttbar$ background than the
standard analysis (which we will refer to as the ``standard cuts'' analysis),
allowing a much higher $\WpWm$ yield for the same number of $\ttbar$ 
events.   Furthermore, the application of a cut on the output of the 
RF classifier hardly alters the distribution of the number of jets in
$\WpWm$~events, in principle allowing for an incisive test of the claim
that the discrepancy between the measured and the predicted values
for $\csWW$ stems from the distribution of the number of jets.

\section{Statement of the Problem}
\par
In the typical hadron collider analysis, the main background
in the $\WpWm$ signal sample comes from $\ttbar$ events.  Our goal is to
study the separation of $\WpWm$ and $\ttbar$ events by classifying them as
effectively as possible.   We are interested mainly in how the RF classifier
makes use of the features of the events; we are not trying to replicate a bona fide
cross section measurement in a high-luminosity collider environment.
For this purpose, we use the \pythia\ event
generator~\cite{PYTHIA}.  We do not attempt to simulate the detector response.
We do, however, limit the fiducial region to $| \eta | < 2.4$ for leptons, and 
$| \eta | < 4.7$ for jets.   There are no inefficiencies and no contamination
in the identified electron, muon, and jet collections (including $b$-tagging).
For these reasons, the performance reported here will be unrealistic
as will be reflected in the efficiencies obtained by applying the standard cuts;
they are more performant than when a realistic simulation is used.
Nonetheless, the relative difference between the standard cuts analysis and the
RF classifier approach should be indicative.
\par
A pre-selection of events is based on the most basic features.
There must be two leptons of opposite charge, and they both must
satisfy $p_T > 20~\GeV$ and $| \eta | < 2.4$.   Jets are reconstructed
with the \pyclus\ routine native to \pythia, for $| \eta | < 4.7$ and $p_T > 15~\GeV$.
Various kinematic quantities are calculated at the level of the \pythia\
event generator and are summarized in Table~\ref{tab:quantities}.
In this context, ``transverse'' means transverse to the beam line, so
$p_T$ refers to the component of a momentum vector that is transverse
to the beam.  The azimuthal angle is denoted by $\phi$.
The missing transverse energy, $\MET$, is calculated as the vector
sum of the momenta of all neutrinos in the event -- detector resolution
effects are not taken into account.  The ``projected'' missing transverse
energy, $\METproj$, takes into account the possible mismeasurement
of lepton energies in Drell-Yan events and is specified in Ref.~\cite{CMSWW}.
A jet that aligns with a $b$~quark is
considered a $b$~jet with no inefficiency and no background.
A kinematic quantity that is absent (for example, the $p^{\mathrm{jet}}_{T3}$ 
for an event with only two jets) is set to~$-99$.
An integer flag, $\LChan$, has value $-2$ for an event with an $e^+e^-$ pair,
$-1$ for a $\mu^+\mu^-$ pair, and $+1$ for an $e^\pm \mu^\mp$ pair.

\begin{table}[h]
\begin{center}
\caption[.]{\label{tab:quantities}
Summary of kinematic quantities used as input features to the random forest}
\begin{tabular}{|cl|}
\hline
$p_{T1}$ & transverse momentum of the leading lepton \\
$p_{T2}$ & transverse momentum of the trailing lepton \\
$q_T$ & transverse momentum of the lepton pair \\
$\MLL$ & di-lepton invariant mass \\
$\Delta\phi$ & difference in azimuthal angle between the leptons \\
$\Nextra$ & number of extra leptons \\
$\pTextra$ & $p_T$ of the leading extra lepton, if any \\
$\Njet$ & number of hadronic jets \\
$\pTjeti$ & $p_T$ of the $i^{\mathrm{th}}$ jet \\
$\Njetb$ & number of jets with $b$-hadrons \\
$\MET$ & missing transverse energy \\
$\METproj$ & ``projected'' missing transverse energy \\
$\Delta\phi_{\ell\ell\mathrm{miss}}$ & angle between the lepton pair and the 
missing momentum vector \\
$\Delta\phi_{\ell\ell j}$ & angle between the lepton pair and
the leading jet \\
$\LChan$ & flag to indicate lepton channel \\
\hline
\end{tabular}
\end{center}
\end{table}

\par
In order to compare to a standard analysis, such as the one developed
by the CMS Collaboration~\cite{CMSWW}, we tried to implement a cut-based
selection of $\WpWm$ events.  The cuts are specified in Table~\ref{tab:cuts}
and are applied to events that pass the pre-selection.

\begin{table}
\begin{center}
\caption[.]{\label{tab:cuts}
Selection criteria for the cut-based analysis}
\begin{tabular}{|c|}
\hline
$\MLL > 12~\GeV$ \\
$q_T < 45~\GeV$ \\
$\Delta\phi_{\ell\ell j} < 2.8798$~rad \\
$\Nextra = 0$ \\
$\Njet = 0$ \\
$\Njetb = 0$ \\
If $\LChan < 0$ then $\MLL < 76~\GeV$ or $\MLL > 106~\GeV$ \\
If $\LChan < 0$ then $\METproj < 20~\GeV$, or
 if $\LChan = 1$ then $\METproj < 45~\GeV$ \\
\hline
\end{tabular}
\end{center}
\end{table}

\section{Method}
\par
For this study, we employed the freeware package \opencv~\cite{OPENCV},
written in {\tt C++} with no reference to external libraries or packages.
We found this package to be fast, effective, and easy to use.
\par
The \pythia\ event generator was run to generate a fixed number of events
from a given process passing the preselection.  For each event, the kinematic
quantities listed in Table~\ref{tab:quantities} were written to a formatted file
and labeled according to the physics process that was simulated:
$K = 1$ for $pp \rightarrow \WpWm$, $K = 2$ for $pp \rightarrow \ttbar$,
and $K = 3$ for the Drell-Yan production of lepton pairs.  
Independent training and testing sets were generated, each consisting 
of $10^4$ $\WpWm$ events, $10^4$ $\ttbar$ events, and $4998$ Drell-Yan events.
Since the number of $\WpWm$ and $\ttbar$ events is the same, the RF
responds only to the features and not to the frequency of events.
The Drell-Yan events are easily eliminated so we do not consider
them further, in this study.
\par
The RF was specified to have a population of 1000 trees. Since the RF does not overfit, its performance improves as the number of trees increases. However, growing the forest is a computationally expensive process whose cost increases with the number of trees, and whose benefit becomes negligible once this number is sufficiently high (over several hundred). A greater number of trees is also beneficial in that it allows for more precision in the output of the classifier, $Y$, which is given by the number of trees categorizing an event as signal divided by the total number of trees. It is therefore quantized at the level of $\Delta Y = 0.001$ in our case.   
\par
Each tree was set to have a maximum depth of 15 nodes, with the number of active features at each node set to the square root of the total number of features. The tree depth is the number of node splits through which a case is processed before the tree assigns it a final classification. A greater depth results in each tree eliminating a smaller number of the training cases at each node, making for a more refined analysis. The tree determines where to split each node from a subset of features, known as the active features, which is selected randomly at each node. The feature supplying the best possible division at each node is the one used to make the split, and choosing that feature from a randomly-selected subset reduces the correlation among trees. In most cases, a subset size near the square root of the total number of features provides the most accurate performance of the~RF. Nonetheless, we tried wide variations of this parameter and the others, and found that they were not critical provided they were not too small.
\par
\begin{figure}
\begin{center}
\includegraphics[width=0.85\textwidth]{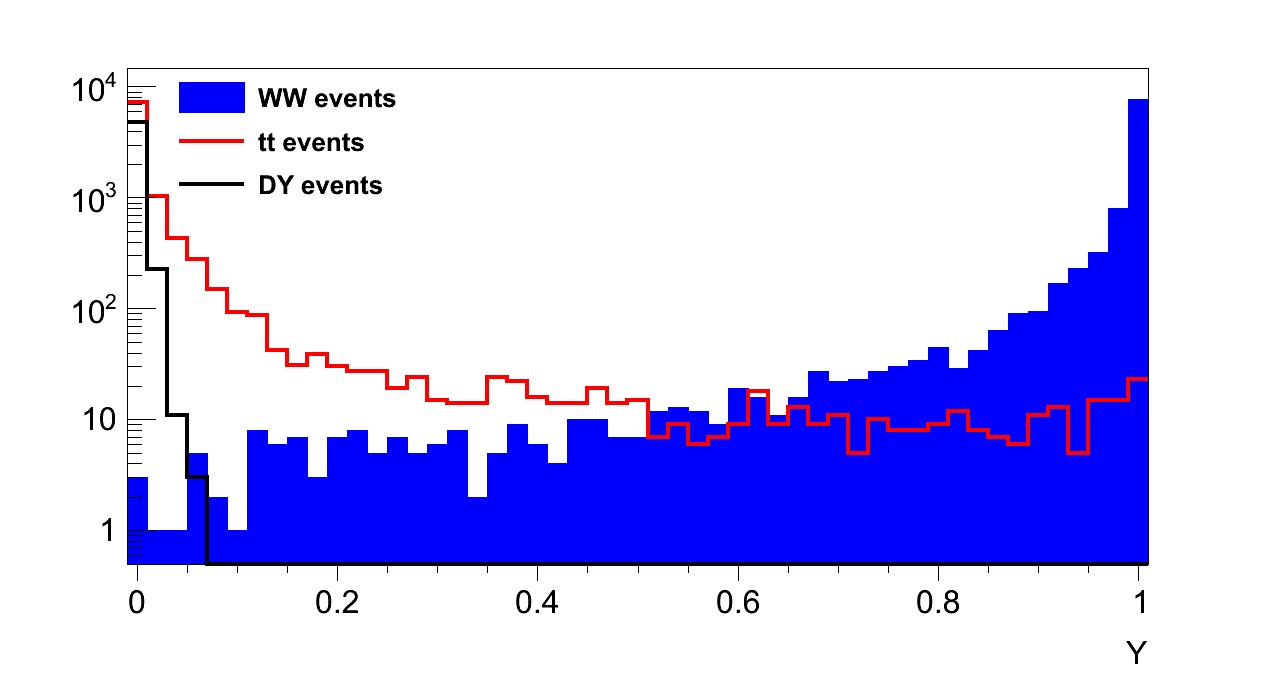}
\caption[.]{\label{fig:Yhist}
Histogram of the output variable $Y$ from the random forest.
The blue histogram is $\WpWm$ signal and the red histogram
is the $\ttbar$ background.   The blue line represents the 
Drell-Yan background.   A cut $Y >  \Ycut$ selects a relatively
pure signal sample.}
\end{center}
\end{figure}

The RF was trained to give an output of $Y = 1$ for signal $\WpWm$ events and
$Y = 0$ for background $\ttbar$ and Drell-Yan events.   After training (which takes
only about a minute), we presented the RF with the test set and made a histogram
of the output~$Y$ -- see Fig.~\ref{fig:Yhist}.  The effective separation of $\WpWm$
signal events from $\ttbar$ and Drell-Yan background events is evident.

\section{Results}
\par
A requirement $Y > \Ycut$ amounts to a selection of signal $\WpWm$ events
and a rejection of $\ttbar$ and Drell-Yan events.   If we vary $\Ycut$ from low to
high values, we can trace out a receiver operator characteristic (ROC) curve.
Following tradition in particle physics, we use the efficiency for $\WpWm$ events,
$\effWW$, and the efficiency for $\ttbar$ events, $\efftt$, to specify the curve.
They are determined using the test set of events; recall that these events have
passed the preselection so $\effWW \rightarrow 1$ means that all events with
a pair of oppositely-charged leptons with $p_T > 20~\GeV$ and $| \eta | < 2.4$
are selected.
\par
\begin{figure}
\begin{center}
\includegraphics[width=0.85\textwidth]{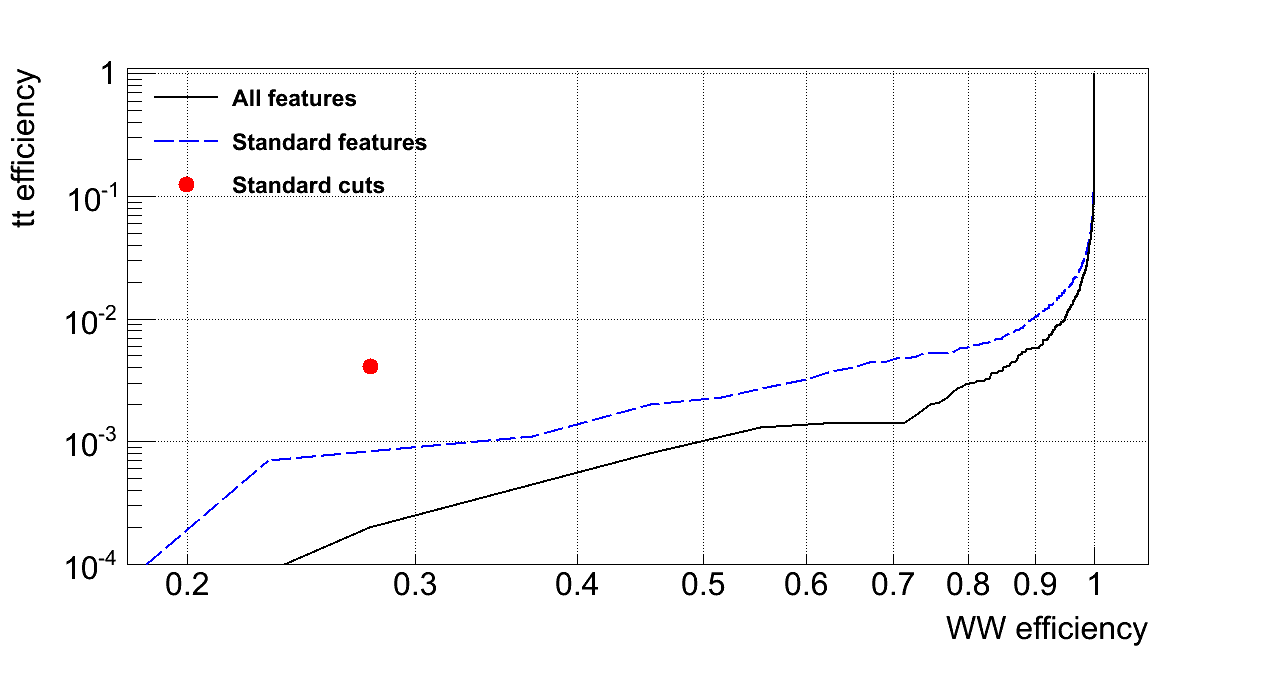}
\includegraphics[width=0.85\textwidth]{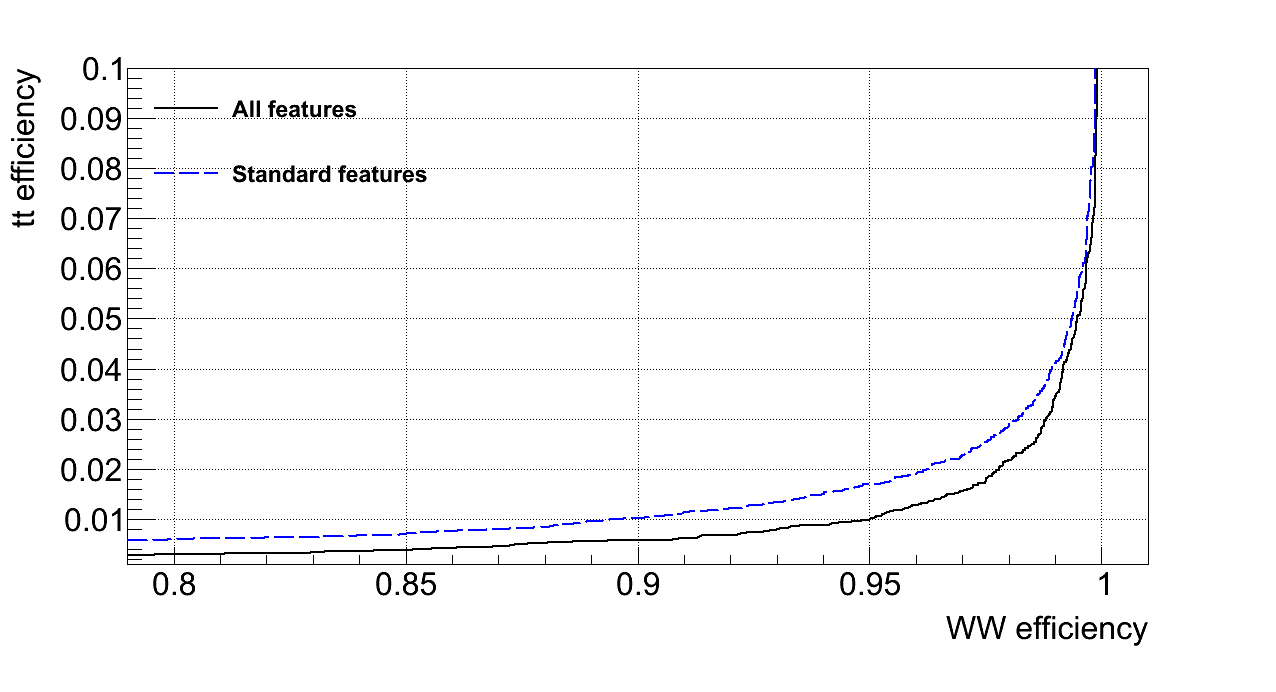}
\caption[.]{\label{fig:ROC}
Receiver operator curves expressed as in terms of the selection efficiency
for signal $\WpWm$ events and background $\ttbar$ events.  The selection
efficiency is defined with respect to events passing the fiducial requirements.
The solid black line is the result obtained with all features listed in
Table~\ref{tab:quantities} while the dashed blue line is the result obtained
only with the features used in the standard cuts analysis (Table~\ref{tab:cuts}).
The performance of the standard cuts is indicated by the red dot in the
upper plot.  The lower plot shows a zoom of the high-$\effWW$ region.}
\end{center}
\end{figure}

Figure~\ref{fig:ROC} shows the ROC curve for the RF.   The jumps and
non-smooth behavior at low values of~$\efftt$ are consequences of the
discretization of $Y$, not of the size of the testing sample.  The fact
that the curve bends sharply toward the lower-right corner of the plane,
where $\effWW$ is high and $\efftt$ is low, reflects the excellent separation
of signal and background events.   The performance of the standard cuts
analysis is represented by the red dot.   Clearly, it is less performant
than the RF.   For a given $\ttbar$ efficiency, the RF achieves a higher
signal efficiency, and likewise, for a given $\WpWm$ efficiency, it achieves
a greater background rejection.   The Drell-Yan background is easily
eliminated as is also the case for the standard cuts analysis: in a realistic
analysis, the Drell-Yan background appears through reconstruction and
detector effects, which are outside the scope of this study.  Note that
in the CMS analysis~\cite{CMSWW}, the primary background is, by far,~$\ttbar$.
\par
We point out that the efficiencies we obtain for the standard cut are
significantly higher than in the actual CMS analysis because our simulation
is idealized.  Presumably, the efficiencies we obtain for the RF are also
``optimistic''  and would not be achieved by a RF trained and evaluated
with a realistic simulation.   Nonetheless, the difference in the performance
of the RF and the standard cuts analysis seen should still carry over to
an analysis wit a realistic simulation of the detector response.
\par
Several quantities are available to the RF that were not used by the standard cuts
analysis.   One might suspect that the improved performance resulted from
the availability of more information.  We trained a second RF in which we
restricted the kinematic quantities to those listed in Table~\ref{tab:cuts}.
The performance is only slightly weaker than the original RF, indicating 
that the RF technique, and not the additional kinematic quantities, 
brings the better performance.

\par
The goal of the LHC analyses is to measure the cross section $\sigmaWW$.
In order to gauge the impact of the RF on this goal, we define a figure of merit
that captures the statistical power of a data set.    In the context of this study,
the cross section can be defined:
\begin{equation}
\label{eq:sigWW}
 \sigmaWW = \frac{\NWW - \Ntt}{\AWW \effWW \Lum}
\end{equation}
where $\AWW$ is the acceptance ({\it i.e.}, the fraction of $\WpWm$ events passing
the preselection), $\Lum$ is the integrated luminosity, 
$\NWW (= \AWW \effWW \Lum \sigmaWW)$ is the number of selected $\WpWm$ signal events,
and $\Ntt ( = \Att \efftt \Lum \sigmatt)$ is the number of selected $\ttbar$ background
events.   Since $\Att \approx \AWW$, the point of the
analysis is to work in a regime in which $\effWW \gg \efftt$.
\par
\begin{figure}
\begin{center}
\includegraphics[width=0.85\textwidth]{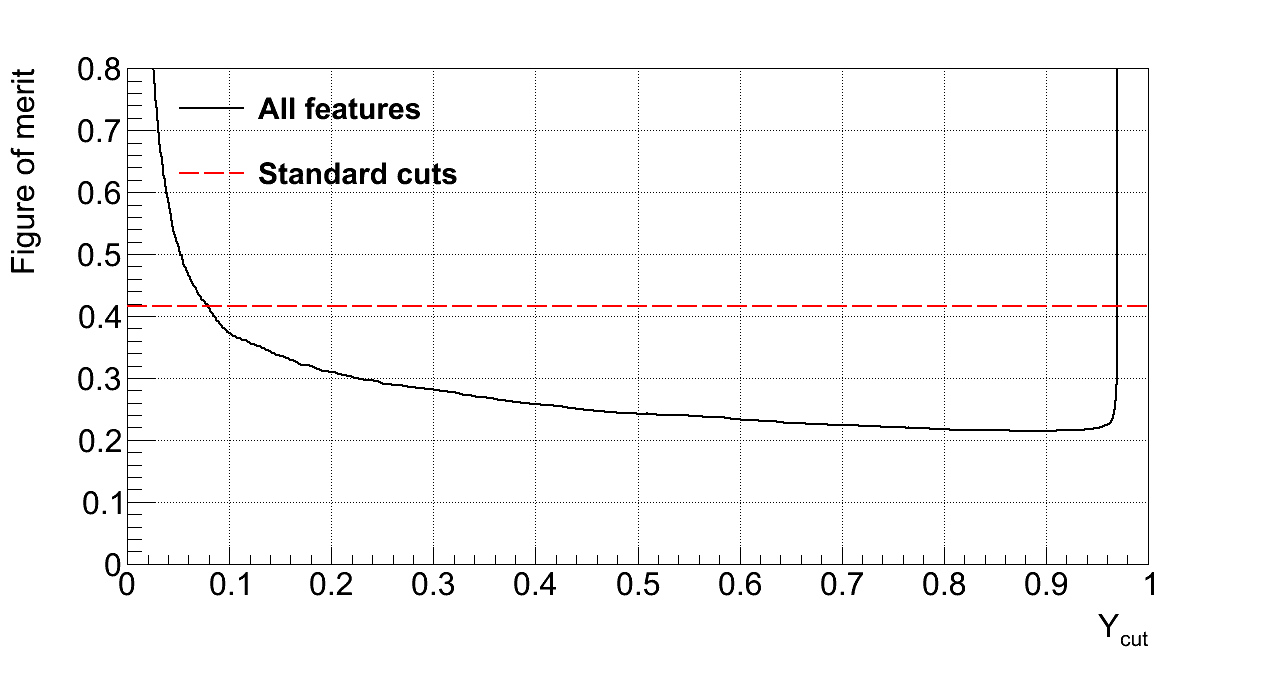}
\caption[.]{\label{fig:fom}
Figure of merit as a function of the cut on the RF output variable, $\Ycut$.}
\end{center}
\end{figure}

We take $A$, $\epsilon$, and $\Lum$ to be sufficiently well known, and focus on the
stochastic quantities~$\NWW$ and~$\Ntt$.   Our figure of merit is the relative statistical
uncertainty on $\sigmaWW$:
\begin{equation}
\label{eq:FOM}
  F  = \frac{\sqrt{\NWW + \Ntt}}{\NWW - \Ntt} 
    = \frac{\sqrt{\AWW \effWW\sigmaWW + \Att\efftt\sigmatt }}{\AWW\effWW\sigmaWW - \Att\efftt\sigmatt} .
\end{equation}
We calculate that  $\AWW = 0.762\,\Att$ based on simulations. The numerical values of
the cross sections at 8~TeV are $\sigmaWW = 57~\pb$~\cite{sigmaww} and 
$\sigmatt = 246~\pb$~\cite{sigmatt}.
Table~\ref{tab:FOM} compares values of~$F$ for the standard cuts analysis
and three operating points for the RF analysis.   The numerical values indicate that the
RF analysis could lead to a statistically more precise measurement than the
standard cuts analysis.   For example, setting $\Ycut$ so that $\efftt$ is the same as
in the standard cuts analysis, the signal efficiency is three times higher
and the corresponding statistical uncertainty would be nearly half as small.
This operating point is near the optimal point defined by the lowest 
value of~$F$.   Figure~\ref{fig:fom} shows the variation of~$F$ with~$\Ycut$;
a shallow minimum is observed near~0.1 reflecting the strong separation 
illustrated in Fig.~\ref{fig:Yhist}.
If, instead of aiming for the best statistical precision, 
we aim for the same statistical uncertainty as
the standard cuts analysis, then the signal efficiency is essentially unity.

\begin{table}
\begin{center}
\caption[.]{\label{tab:FOM}
Comparison of the figure of merit ($F$) for the standard cuts analysis
and three RF operating points}
\begin{tabular}{|lcccc|}
\hline
analysis & $\Ycut$ & $\effWW$ & $\efftt$ & $F$ \\
\hline
standard cuts          & - & $0.2767$ & $0.0041$ &  $0.417$  \\
RF low-$\efftt$        & $0.965$ & $0.8576$  & $0.0041$ & $0.217$  \\
RF best-$F$            & $0.914$ & $0.9061$ & $0.0058$ & $0.214$ \\
RF high-$\effWW$ & $0.072$ & $0.9987$ & $0.0772$  & $0.419$ \\
\hline
\end{tabular}
\end{center}
\end{table}

\par
The crucial piece of the standard cuts analysis is the so-called ``jet veto.''
As mentioned earlier, it has been suggested that the Monte Carlo event simulators in use by the
LHC collaborations are not sufficiently accurate, and that the discrepancy
between the measured values of $\sigmaWW$ and the predicted value
is caused by an incorrect value of $\effWW$ stemming from the jet veto.
\par
An intriguing feature of the RF selection is that it does not result in
strong cuts in any variable.   Some trees may make use of a given variable
while others will not; the result is a softer impact on the original distribution
of that variable.  
\par
\begin{figure}
\begin{center}
\includegraphics[width=0.85\textwidth]{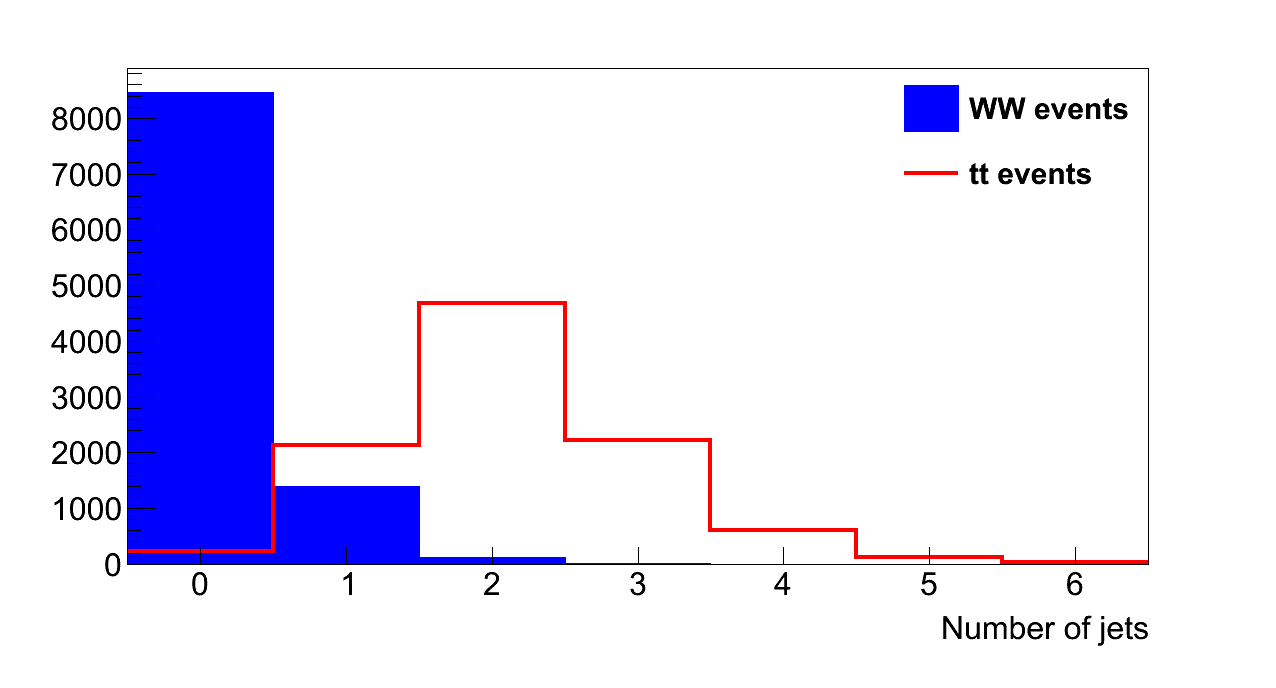}
\includegraphics[width=0.85\textwidth]{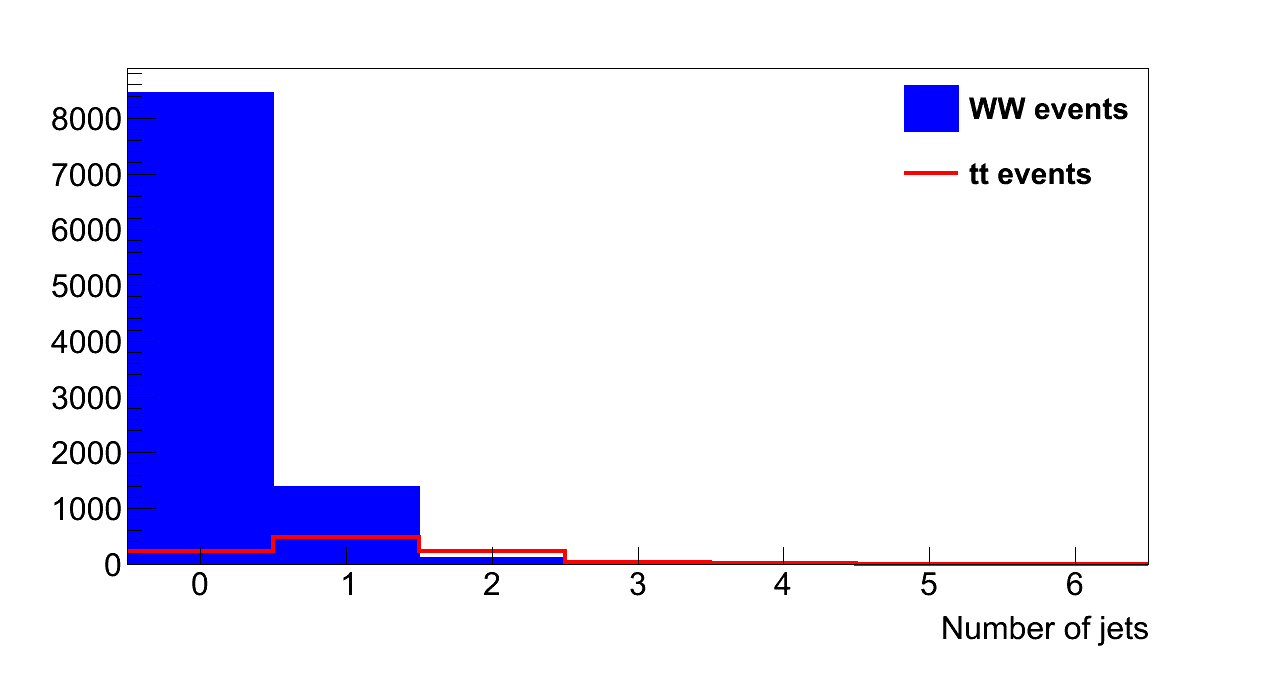}
\caption[.]{\label{fig:njet}
Distribution of the number of jets, $\Njet$.   The solid blue histogram
represents the $\WpWm$ distribution, and the thick red line represents
the $\ttbar$ distribution.   The top plot shows the distributions after the
pre-selection (there are $10^4$ entries for each), and the bottom plot
shows the distributions for the high-$\effWW$ working point listed
in Table~\ref{tab:FOM}.   The $\WpWm$ distribution is essentially unchanged
despite the strong suppression of $\ttbar$ events by a factor of nine.}
\end{center}
\end{figure}

We studied the $\Njet$ distribution before and after applying a selection
cut on the RF output.  Remarkably, there was {\em no impact on $\Njet$ at all}
if the cut was not too strong.   For example, if we chose to match the value
of $\efftt$ obtained with the standard cuts analysis, then with the RF the value
of $\effWW$ is very high and the $\Njet$ distribution is unchanged, as illustrated
in Fig.~\ref{fig:njet}.  This feature of the RF would allow, in principle, a direct
check of the $\Njet$ distribution in $pp \rightarrow \WpWm$ events, and
innocuate the cross section measurement from  any systematic bias coming from the jet veto.
In a more general sense, the fact that the RF classifier induces relatively 
gentle distortions of the distributions of key kinematic quantities (certainly
less abrupt than applying a cut) suggests that systematic uncertainties
due to the modeling of these distributions should be significantly reduced
or perhaps effectively eliminated.

\par
A common concern with multi-variate methods is the potential sensitivity
to the training set.  In a real analysis, one would use simulated events to
train the RF, and then apply the RF to real events recorded with the detector.
If the fidelity of the simulation is imperfect, then a bias may result that is
difficult to quantify.   Since we do not simulate the detector at all, we cannot
address this question carefully.   Nonetheless, we did apply a gross distortion
to the $\MET$, rescaling it upward by 10\% in the test sample (but not in the
training sample).   The change in the efficiencies are $\Delta\effWW = -0.05$
and $\Delta \efftt = 0.002$; the figure of merit changed  by only $\Delta F = 0.006$.
\par
Finally, we checked one of the main virtues of the RF classifier: it should be
able to classify an event even if it is missing one or more features.
For example, one might disregard the $\MET$ for an event (and of course
the associated variables) if there was evidence for anomalous noise
in the calorimetry.  For a standard analysis, the event must be discarded; one
usually restricts the analysis only to those data sets for which the entire
detector was operating well.   If ``bad'' or otherwise faulty data could be
recovered using an RF, then a significant gain in the size of the data set
could be realized.   We made a test in which we masked the $\MET$ for
a random sample of 10\% of the testing set (but not in the training set).
Remarkably, the loss in $\effWW$ was only~$2.4\%$.

\section{Conclusions}
\par
We have carried out am exercise in the application of random forest classifiers to
separating signal and background events in particle physics.  A technically
similar exercise was performed by members of the MAGIC Collaboration,
who were separating photon- and hadron-initiated atmospheric showers~\cite{MAGIC}.
Other applications of random forests in high-energy physics are documented
in~\cite{D0top,D0VV,D0Higgs,D0bjet,BaBarbsgamma,BaBarHiggs}.
\par
For this exercise, we produced training and testing samples at the parton level,
and presented them to a random forest.   The RF performance surpasses the
performance of a standard cut-based analysis.  Furthermore, the distortion of
the distributions of key kinematic event features is relatively slight, suggesting that
systematic uncertainties due to modeling might be reduced.   Finally, the RF
we developed can tolerate missing features such as the missing transverse
energy without a severe degradation of its performance, a characteristic that
may allow less than perfect data to be utilized in measurements and searches
for new particle and phenomena.

\section*{Appendix: $\sigmaWW$ Measurements}
\par
The measured cross sections for $pp \rightarrow \WpWm$ are higher than
predicted at the level of two or more standard deviations.  Table~\ref{tab:sigmaWW}
gives a summary.  The CMS and ATLAS collaborations have made two measurements
each: one at $\sqrt{s} = 7$~TeV and the other at 8~TeV.   The CMS and ATLAS values
are mutually consistent, and are higher than the theoretical predictions.  For the 
calculations of the ratios listed in the table, the asymmetric uncertainties have been
symmetrized.

\begin{table}[hb]
\begin{center}
\caption[.]{\label{tab:sigmaWW}
Summary of measurements of $\sigmaWW$ in pb.
For the measured values, the first uncertainty is statistical,
the second is systematic, and the third comes from the luminosity.
The last column shows the ratio of the experimental value and
the theoretical one.}
\begin{tabular}{|lc|cc|c|}
\hline
Collaboration & $\sqrt{s}$ (TeV) & measured & theoretical & ratio\\
\hline
ATLAS \cite{ATLASWW} (prelim.) & 8 & $71.4 \pm 1.2~^{+5.0}_{-4.4}~^{+2.2}_{-2.1}$ & $58.7^{+3.0}_{-2.7}$ 
 & $1.22 \pm 0.11$ \\
CMS \cite{CMSWW} & 8 & $69.9 \pm 2.8 \pm 5.6 \pm 3.1$ & $57.3^{+2.3}_{-1.6}$ 
 & $1.22 \pm 0.13$ \\
ATLAS \cite{ATLASWW7} & 7 & $51.9 \pm 2.0 \pm 3.9 \pm 2.0$ & $44.7^{+2.1}_{-1.9}$ 
 & $1.16 \pm 0.12$ \\
CMS \cite{CMSWW7} & 7 & $52.4 \pm 2.0 \pm 4.5 \pm 1.2$ & $47.0 \pm 2.0$ 
 & $1.12 \pm 0.12$ \\
\hline
\end{tabular}
\end{center}
\end{table}



\begin{thebibliography}{12}
\bibitem{sigmaww}
This value was obtained using MCFM.
\end{thebibliography}
\end{document}